\documentclass[a4paper,superscriptaddress,aps,prl,twocolumn,floatfix,citeautoscript,showpacs,floatfix]{revtex4-1}

\usepackage{graphicx}
\usepackage{epsfig}
\usepackage{color}
\usepackage{latexsym}
\usepackage{amsmath,bm}
\usepackage{natbib}
\newcommand{\lsi}{Laboratoire des Solides Irradi\'es and ETSF, \'Ecole Polytechnique,
CNRS, CEA-DSM, 91128 Palaiseau, France}
\newcommand{\lpmcn}{LPMCN, Universit\'e Claude Bernard Lyon I and CNRS, 69622
Villeurbanne, France}
\begin{document}

\title{Excitonic effects in the optical properties of CdSe nanowires}

% A known bug in revtex4.1 Using the same \affiliation more than once results in some authors not getting any affiliation at all.
 \author{Jos\'e G. Vilhena$^{1}$, Silvana Botti$^{2,1}$ and Miguel A. L. Marques$^1$}
 \affiliation{$^{1)}$\lpmcn \\ $^{2)}$\lsi}

\date{\today}

\begin{abstract}
Using a first-principle approach beyond density functional theory
we calculate the electronic and optical properties of small diameter CdSe nanowires.
Our results demonstrate how some approximations commonly used
in bulk systems fail at this nano-scale level and how indispensable it is to
include crystal local fields and excitonic effects to predict the
unique optical properties of nanowires. From our results, we then construct a simple
model that describes the optical gap as a function of the diameter of the wire, that turns out to be in
excellent agreement with experiments for intermediate and large
diameters.
\end{abstract}

\pacs{}

\maketitle

Nanowires (NW) exhibit a wide range of unique
properties~\cite{intro-05.Review}, including tunable band
gaps, ballistic transport~\cite{intro-09},
optical anisotropy~\cite{CdSe-exp01} and strong excitonic
effects~\cite{intro-08.Nature}. It is therefore not surprising that for
the past twenty years NW have emerged as one of the most active fields
of research in material science~\cite{intro-05.Review}. This growing
interest is mainly due to their short term promising
applications~\cite{intro-09,intro-06,intro-08.Nature} and is empowered
by a strong demand from industries for smaller and more effective
devices. To a large extent, the novel properties
emerge from the lateral confinement of the electrons in the wire,
leading to the blue shift of the electronic band gap with decreasing
diameter.

In this article we focus on the optical properties of CdSe NW
from a first-principle perspective, based and going
beyond standard density functional theory (DFT). Beside analyzing how
optical properties evolve with the diameter of the wire, we
test some approximations that are commonly used for bulk materials
and discuss their applicability to electronic excitations in 1D
systems.
Several studies of the optical properties of CdSe wires have appeared in the past
years~\cite{CdSe-theo.semiempiricalo,cdse-theo1,CdSe-theo.semiempiricalF}. Classical~\cite{CdSe-exp01} and
semi-empirical~\cite{CdSe-theo.semiempiricalF} methods have been
quite successful in describing large-diameter wires, but they fail for small and
medium diameters. Furthermore, none of articles using first-principle
methods present in the literature~\cite{CdSe-theo.semiempiricalo,cdse-theo1} was capable of
capturing the physics of electronic excitations in these confined systems.
This was due to the neglect of at least one of the following physical effects that
are fundamental for an accurate description of nano-scale
objects:

(a) Crystal local-field effects. When the polarization of the light is
perpendicular to the long axis there is an accumulation of
charges at the wire surface, which in turn is responsible for the
attenuation of the electric field inside the NW. This leads to a
strong suppression of the absorption for light polarized perpendicular
to the NW axis, increasing dramatically the optical anisotropy of the
system~\cite{Marinopoulos2003, *Bruneval2005-comm}. This huge effect has already been measured~\cite{CdSe-exp01}
in polarized photo-luminescence experiments, but unfortunately some
recent theoretical works~\cite{cdse-theo1, SiNW} still neglect it.

(b) Electron-electron interaction and excitonic effects. 
These effects are very pronounced in
semi-conducting NW due to the attenuation of the screening. The strong electron-hole binding
is responsible for a red shift of the fundamental
absorption frequency. Due to this effect, NW can be seen as exciton
traps, a property that endows them of great technological
interest~\cite{intro-08.Nature}.

\begin{figure}[t]
  \centering \includegraphics[width=0.9\columnwidth]{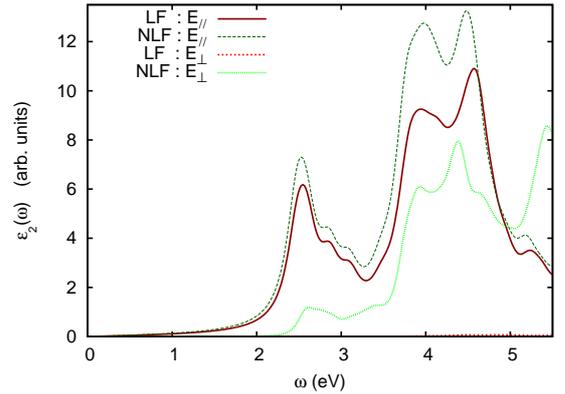}
  \caption{(Color online) Imaginary part of the dielectric function
  for the smallest NW ($d=0.73$\,nm), calculated in the RPA
  with (LF) and without (NLF) crystal local fields, for light polarized
  parallel ($E_{\parallel}$) and perpendicular ($E_{\perp}$) to the wire axis.
  Note that the absorption is zero for NLF:$E_{\perp}$.}
  \label{fig:fig1}
\end{figure}

To study the importance of local-field corrections, we calculated the absorption spectra
of unpassivated CdSe NW of 5 different diameters (0.73, 1.17, 1.59,
1.77, and 2.01\,nm) starting from DFT Kohn-Sham states and applying the random-phase approximation (RPA)
to obtain the dielectric tensor~(for details on this and all the other approximations employed in
the following, refer to Ref.~\onlinecite{botti-review-07}). The neglect of local-field effects is equivalent to
applying Fermi's golden rule, i.e., to treat only independent particle transitions, and completely ignores 
the inhomogeneity in the dielectric response due to the reduced dimensionality of the nano-object.

All CdSe NW were assumed to be infinitely long with periodic length 7.01\,\AA\, and
with their axis parallel to the wurtzite (001) axis. Ground state
calculations were performed using the DFT code
ABINIT~\cite{Software-ABINIT1}, and the core
electrons of Cd ([Kr]4d10) and Se ([Ar]3d10) were described by Hamann
norm conserving pseudo-potentials. We chose the
Perdew-Burke-Ernzerhof~\cite{PBE} approximation to the
exchange-correlation functional. Converged calculations required
a cutoff energy of 20 Ha and a $1\times1\times8$ Monkhorst-pack
sampling of the Brillouin zone. Atomic positions were relaxed starting 
from the bulk wurtzite structure. The converged spacing between the wires in our
supercell approach was at least 7\,\AA. 

The optical spectra of the wire with 0.73\,nm, calculated at the RPA
level including or neglecting local fields is shown in
Fig.~\ref{fig:fig1} .  Without local fields, there is a small
anisotropy between the absorption perpendicular and parallel to the NW
axis, comparable to the anisotropy in bulk CdSe.
However, and as expected, turning on local-field effects suppresses
completely the low energy absorption peaks in the perpendicular
direction, rendering the wire almost transparent below 6.5\,eV. This
optical anisotropy, in agreement with experimental
results~\cite{CdSe-exp01}, clearly decreases with the diameter, but it
is known to be still relevant for NW with a diameter of 100\,nm.

\begin{figure}[t]
  \centering \includegraphics[width=0.9\columnwidth]{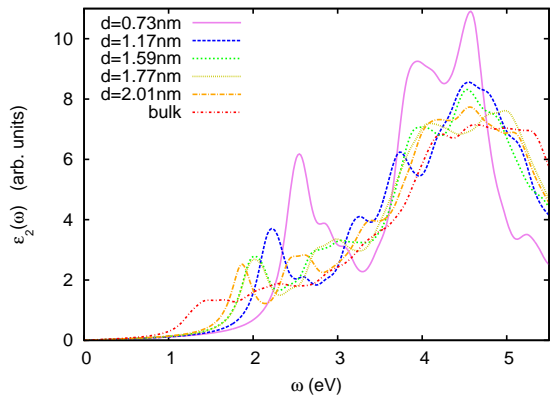}
  \caption{(Color online) Dependence of the RPA absorption spectrum
  with the diameter of the NW. For comparison we also show
  the RPA result for the wurtzite CdSe bulk material.}
  \label{fig:fig2}
\end{figure}

To observe how the optical properties vary with
diameter, we plot the RPA spectrum (including local fields) for 
five small wires in Fig.~\ref{fig:fig2} . It is expected that the RPA gives a
poor quantitative result due to the inappropriate treatment of electron-electron 
and electron-hole interactions, but it is however possible to extract
qualitative trends thanks to the partial cancellation of these two terms (see below). We see that increasing the diameter, and thereby
decreasing the confinement effect, leads to a red-shift of the
spectrum, with the absorption threshold moving towards the RPA bulk
value. There is also a redistribution of the oscillator strengths,
with the first peak loosing intensity.

\begin{figure}[t]
  \centering
  \includegraphics[width=0.9\columnwidth]{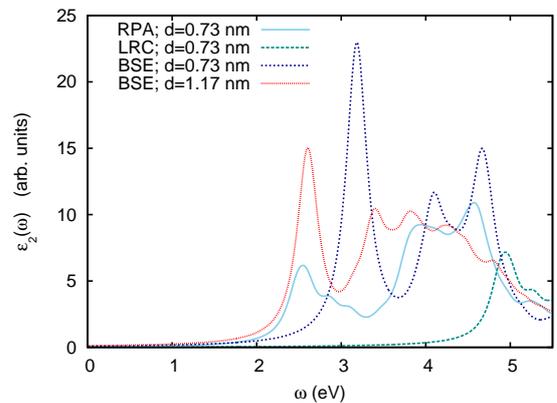}
  \caption{(Color online) Imaginary part of the dielectric constant in
  the different approximations: RPA, LRC kernel, and BSE for the $d=0.73$\,nm NW. The
  ALDA results (not shown) are almost identical to the
  RPA. Also shown is the BSE results for the $d=1.17$\,nm NW.
  } 
  \label{fig:fig3}
\end{figure}

To go beyond the RPA and to include the relevant missing contributions, we
performed calculations solving the Bethe-Salpeter equation (BSE)~\cite{Onida2002} and 
using time-dependent DFT~\cite{botti-review-07}
--- both in the adiabatic local density approximation (ALDA) and with the
model long range (LRC) kernels~\cite{Other-LRCab, *Other-LRCa} derived from the BSE.
The BSE approach is the state-of-the-art method for calculations of optical absorption
and gives results in excellent agreement with experiments~\cite{Onida2002,botti-review-07}.
For the LRC and BSE calculations, one requires as a starting point the quasi-particle
band structure, usually obtained within the GW approximation~\cite{Hedin1965}. Note
that the GW approximation predicts accurate band gap energies for CdSe, in contrast
with the systematic underestimation of the gap obtained in DFT.
Unfortunately, GW calculations are computationally
demanding, even for the small wires studied here. We therefore
used the following approach to obtain the quasi-particle corrections:
we solved the GW equations for our smallest NW ($d=0.73$\,nm), and
then used a simple theoretical model~\cite{Other-Model} based on
crystal field theory to interpolate between this result and the GW gap
of the bulk~\cite{cdse-theo4.Angel}.

The GW calculations were performed with ABINIT, applying the standard
plasmon-pole approximation. An energy cutoff for the dielectric matrix
of 9\,Ha and the technique of
Ref.~\onlinecite{Other-Bruneval.trick} to reduce the number of
unoccupied states in the sums over states were employed. Despite
the use of a cylindrical cutoff for the Coulomb interaction~\cite{Other-Beigi}, a
distance of 10\,\AA\ between the wires was required to converge the GW
corrections to the Kohn-Sham band structure. These corrections turned out to be quite insensitive to
the ${\bf k}$-point and band index, in agreement with other results for
Si and Ge NW~\cite{Other-Si.NW, Other-Ge.NW}, and in contrast
with findings for graphene nanoribbons~\cite{Other-Nribons.S.louie}.
The GW gap for the smallest wire was $E_{\rm gap}^{\rm GW}=4.84$\,eV,
and the final form of the model relating the quasi-particle gap with
the diameter of the wire ($d$ in nm) turned out to be $E^{\rm model}_{\rm GW
gap}(d) = 1.91 + 2.14/d$\,(eV).

The BSE results (see Fig.~\ref{fig:fig3}), obtained using the
code YAMBO~\cite{Software-YAMBO}, prove the existence of strong
excitonic effects, with the excitonic peaks in the visible energy
range. The excitonic binding energy compensates almost entirely the large blue-shift
coming from the quasi-particle corrections, and leads to a transfer of
the oscillator strength from the higher energy absorption peaks to the
first peak. On the other hand, the LRC kernels derived from the BSE, that yield
results comparable to the solution of the BSE for bulk CdSe~\cite{Other-LRCab},
fail dramatically for NW --- in fact, as shown in Fig.~\ref{fig:fig3} ,
even the optical gap resulting from DFT+RPA calculation is in better
agreement with the BSE result. The
attempt to increase the excitonic effect by modulating the empirical 
parameters of the LRC kernel did not lead to improvements. The reason
is that the simple LRC approximations are only valid for delocalized excitons,
but fail to reproduce the considerable binding energy of
the very localized excitons existing in NW.

The exciton binding energies from BSE calculations, for the NW with
diameters of 0.73 and 1.17\,nm are respectively 1.6 and 1.13\,eV. These
values are much larger than the binding energy in bulk
CdSe~\cite{intro-08.Nature}, 5.1\,meV, and almost twice as large as
the ones found in carbon nanotubes ~\cite{nanotubes}. The decrease of the
binding energy with the NW diameter, also followed by a transfer of the intensity
of the absorption-edge peak to the higher energy ones (see
Fig.~\ref{fig:fig3}), reflects the weakening of the excitonic effects
with the increasing of the exciton spatial extent. In fact, the exciton radius of
bulk CdSe is 5.6\,nm~\cite{intro-08.Nature}: A nano-object where one
of its dimensions is smaller than this value will have strongly bound
excitons, due to a large overlap between electron and hole.  Furthermore one expects that the larger the
confinement the larger is the excitonic binding energy.  In our BSE
results we also found spin singlet dark excitons throughout the
spectra, with the first one appearing at 3.2\,eV for the smallest
wire. This dark exciton may play a role in non-radiative decay
processes and thus affect the luminescence properties.

\begin{figure}[t]
  \centering \includegraphics[width=0.9\columnwidth]{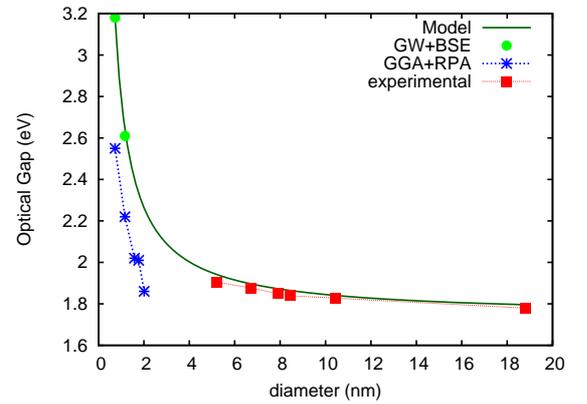}
  \caption{(Color online) Optical gaps calculated with the RPA and
  the BSE compared to experiment~\cite{CdSe-exp02}. The lines connecting the RPA and the
  experimental points are a guide to the eye. The green line is 
  the model interpolation between the smallest NW and the bulk.
  }
  \label{fig:fig4}
\end{figure}

Based on our BSE results for the two smallest wires and on the
experimental optical gap of the bulk~\cite{CdSe-exp02}, we constructed
a simple model to interpolate the optical gaps of CdSe NW. The form is
the same as the previous model for the GW gaps, and the formula
connecting optical gaps and the diameter of the wire turns out to be
$E^{\rm model}_{\rm BSE gap}(d \,\, \textrm{in nm}) = 1.74 + 1.05/d$\,(eV). We can see,
from Fig.~\ref{fig:fig4} , that this model describes extremely well the
experimental results~\cite{CdSe-exp02} not only for 
large NW but also in the intermediate regime. Finally, from
Fig.~\ref{fig:fig4} we conclude that although the RPA results are
red-shifted with respect to the latter they still conserve
approximately the correct slope.

In conclusion, the first-principle calculations
reported here show how indispensable is the inclusion of local-field and excitonic effect
to describe quantitatively the optical response of CdSe NW.
In fact, calculations based on the Fermi's golden rule will never be able to catch the physics
arising from the low dimensionality of the system, which is at the heart of the novel properties
of NW. We have then provided a simple model for the dependency of the optical gap with diameter, which is in excellent agreement with available experimental
results. Furthermore we have observed the failure of long-range model
kernels of time-dependent density functional theory for these nano-scale systems, 
in spite of their success in the calculation of the optical properties of bulk CdSe.

SB acknowledges support from EU’s 7th Framework Programme (e-I3
contract ETSF), MALM from the Portuguese
FCT (PTDC/FIS/73578/2006) and the French ANR
(ANR-08-CEXC8-008-01), JGV from the
FCT (SFRH/BD/38340/2007). Calculations were performed at the LCA of the University
of Coimbra and at GENCI (project x2009096017).

% \bibliography{biblio.bib}

%Merlin.mbs v4.21 2009-07-09.
%

\end{document}